\documentstyle[12pt]{article}
\advance \topmargin by -\headheight
\advance \topmargin by -\headsep

\evensidemargin \oddsidemargin
\begin{document}

\topmargin -.6in
\def\rf#1{(\ref{eq:#1})}
\def\lab#1{\label{eq:#1}}
\begin{titlepage}
\vspace*{-1cm}
\noindent
\hfill{BGU-94 / 17 / June- PH}\\
\phantom{bla}
\hfill{gr-qc/9408031}\\
\begin{center}
{\large\bf Kaluza -- Klein Quantum Cosmology with \\
Primordial Negative Cosmological Constant}\footnotemark
\footnotetext{Talk given at the
School-Seminar on  Multidimensional Gravity and Cosmology,
Yaroslavl, Russia,\\ 20-26 June, 1994. }
\vskip .3in

\noindent
E. I. Guendelman and A. B. Kaganovich

\vspace{18pt}

\begin{small}
  {\it Ben Gurion University of the Negev, Physics Department,\\
 Beer Sheva 84105, Israel\\
(e-mails: ~guendel@bguvm.bgu.ac.il ~and~alexk@bguvm.bgu.ac.il)}
\end{small}
\end{center}
\par \vskip .3in

\begin{abstract}
In many interesting models, including superstring theories,
a negative vacuum energy is predicted. Although this effect is usually
regarded as undesirable from a cosmological point of view, we show that
this can be the basis for a new approach to the cosmology of the early
Universe. In the framework of quantum cosmology (in higher dimensions)
when we consider a negative cosmological constant and matter that could
be dust or, alternatively, coherent excitations of a scalar field, the
role of cosmic time can be understood. Then we can predict the
existence of a ``quantum inflationary phase'' for some
dimensions and a simultaneous ``quantum deflationary phase'' for the
remaining dimensions. We discuss how it may be possible to exit from this
inflation-compactification era to a phase with zero cosmological constant which
allows a classical description at late times.

\end{abstract}

\end{titlepage}

Most standard inflationary models \cite{guth} are based on the assumption of
the existence of a positive cosmological constant, which is the source of the
inflation in the very early Universe. In this process all the spacial
dimensions are  asymptotically exponentially  enlarged. If however, we want to
use as a framework of unification of all forces of nature, a higher dimensional
Kaluza-Klein model of which the superstring alternative is now favored, then
the idea of exponentially expanding all dimensions in the early Universe ceases
to be the most interesting possibility. A more attractive alternative would be
to link dynamically the smallness of the extra dimensions and the big size of
the visible dimensions \cite{venus}. It appears however that in a classical
framework, this idea seems difficult to implement \cite{dima}. Therefore it is
natural to study the possibility of these effects in the context of quantum
cosmology \cite{qic1}.

We have studied \cite{qic1} $1+D$-dimensional, toroidally
compact Kaluza-Klein cosmology where the geometry is defined by
\begin{equation}
ds^2 = - dt^2 + \sum_{j=1}^D a_i^2(t) (dx^i)^2, \quad
0 \leq x^i  \leq 1 \nonumber
\end{equation}
A  negative cosmological constant $\Lambda$ and dust (that is, a
pressureless perfect fluid state) are the sources of gravity which is
described by the Einstein action. The difference between the cases of
negative and positive cosmological constants is that while a positive
cosmological constant generates
an upside-down potential for the volume $V \equiv a_1a_2 \cdots a_D$ of the
Universe, which therefore leads to exponentially increasing volume solutions, a
negative cosmological constant, in contrast, generates a potential which does
not allow the volume of the Universe to expand to very large values.

           In order to demonstrate the properties of the model with
$\Lambda < 0$ it is convenient to choose units where $16 \pi G = 1$ and
make use of the new variables
\begin{equation}
\rho^2= 4(D-1)V/D,\quad \theta_i = \ln (a_i /V^{1/D}), \quad i=1,\ldots ,D
\nonumber
\end{equation}
Notice that  $\theta_1+\cdots + \theta_D \equiv 0$ . The independent
variables that diagonalize the Lagrangian and Hamiltonian are $\rho$ and
\begin{equation}
z^j=\frac{1}{\sqrt{D}+1}\sqrt{\frac{D}{2(D-1)}}\lbrack
\theta_1 + \cdots + (\sqrt{D}+2)\theta_j + \cdots + \theta\;,\;
j=1,\ldots ,D-1 \nonumber
\end{equation}
         The $00$-component of Einstein's equations is the constraint
equation which tells us that the Hamiltonian $H$ of the system is equal to
$0 $ (below $\mu$ denotes the dust density times $V$):
\begin{equation}
H  = \dot{\rho}^2  -  \rho^2
\sum_{j=1}^{D-1} (\dot{z}^j)^2 + \omega^2 \rho^2 - \mu = 0,\quad
\omega^2 = - \frac{D\Lambda}{2(D+1)}> 0
\label{eq:1}
\end{equation}
Here the dot  denotes derivative with respect to cosmic time $t$.
 Due to the symmetry  $z^i\longrightarrow z^i +c^i$
($c^i =$ constants), the quantities $F_i = - 2\rho^2 \dot{z}^i$
($i=1,\ldots ,D-1$) are conserved ones.  Inserting $\dot{z}^i =
-\frac{F_i}{2\rho^2}$ in eq. (\ref{eq:1}) we see that in
classical case the effective potential forces the volume to collapse to zero.
Such singularities may be avoided in quantum cosmology \cite{early} and in our
model, for a large class of operator orderings, the amplitude for zero volume
is exactly zero \cite{qic1}. Together with our choice of  $\Lambda < 0$, this
means that $<V>$ must be non zero and finite.

In quantum mechanics eq. (\ref{eq:1}) becomes $H \Psi = 0$,
the so called
``Wheeler-DeWitt'' (WDW) equation.
The quantized degrees of freedom $(\rho, z^j)$ constitute a
$D$-dimensional minisuperspace. The exact solutions of the WDW equation have
been explicitly found in \cite{qic1}. It was found there that physically
satisfactory
solutions are possible provided the dust content $\mu$ is big enough. Then the
possible values of $F^2\equiv F_1^2+F_2^2 + \cdots + F_{D-1}^2 $ are quantized:
$F^2= (F^2)_n$ here $n$ is a non negative integer (for an explicit expression
of $(F^2)_n$ see \cite{qic1}).

More information concerning the evolution of the Universe in this model can
be obtained from the Heisenberg equations of motion, for example:
\begin{equation}
 \frac{dz^j}{dt} =  i\lbrack H\, ,\,z^j \rbrack  =
\frac{i}{2\rho^2}  \frac{\partial}{\partial z^j}
\label{eq:2}
\end{equation}
Since $z^j$ dependence of $\Psi$ can be taken to be of the form
$\exp (i \sum_{j=1}^{D-1}F_j z^j)$ (due to the symmetry
$z^i\longrightarrow z^i +c^i$), we get that
$-i \frac{\partial}{\partial z^j}\rightarrow F_j$ and taking expectation value
of both sides of eq. (\ref{eq:2}), we get
\begin{equation}
\frac{d<z^j>}{dt}  = - <\frac{1}{2\rho^2}>F_j \nonumber
\end{equation}
which is in agreement with the classical result. Since
one can check that
\begin{equation}
\frac{d}{dt}<\frac{1}{2\rho^2}>=0, \nonumber
\end{equation}
we obtain
\begin{equation}
<z^j>=-
<\frac{1}{2\rho^2}>F_j t + constant,\quad j=1,\ldots ,D-1 \nonumber
\end{equation}
Therefore
\begin{equation}
<\theta_i> = \alpha_{i}t+\gamma_i \quad   i=1,\ldots ,D\;,\;\; {\rm where}
{}~~\alpha_i, \gamma_i \; {\rm ~are~constants~and} \quad
\sum_{j=1}^D \alpha_i = 0
\label{eq:3}
\end{equation}
In the interesting particular case when
$\alpha_1 = \alpha_2 = \alpha_1 \equiv \alpha$
(isotropic evolution of three dimensions)
and $\alpha_4 = \ldots = \alpha_D \equiv \tilde{\alpha}$
(isotropic extra dimensional evolution), we get \cite{qic1}:
\begin{equation}
\tilde{\alpha}=-\frac{3\alpha}{D-3}, \quad
\alpha =\pm \frac{2\omega^2}{D\lbrack \mu-2(n + \frac{1}{2})|\omega|
\rbrack \sqrt{\frac{(D-3)(D-1)}{3}}} | F|_n \;, \quad
|F|_n\equiv \sqrt{(F^2)_n} \nonumber
\end{equation}
Notice that in spite of the fact that the Hamiltonian acting on the wave
function vanishes, cosmic time dependence of the expectation values of the
variables $\theta^i$ appears. This seems to be strange because formally
$<\frac{dQ}{dt}>= i <\lbrack H, Q \rbrack >= 0$  (since $H| > = 0$ and
$< |H = 0$) for any not explicitly dependent on time observable Q.
This is a way of
formulating the well known problem of non appearance of time in quantum
cosmology \cite{kuchar}. Appearance of cosmic time dependence in
eq. (\ref{eq:3}) however
does not seem so strange if one notices that the wave function dependence on
$z^j$ and also Heisenberg equations (\ref{eq:2}) are very similar to those
of a free non
relativistic particle for which a linear time evolution in the average position
is also obtained. The formal argument that gave us $<\frac{dQ}{dt} > = 0$
fails \cite{qic1} for  $Q = z^j$ because $H$ fails to be hermitian for badly
behaved states as  $|z^j|$ approches infinity.

           We see that for this model  {\em it is possible to describe
cosmological evolution (proceeding in cosmic time) in terms of averages of
quantum cosmology variables}. The quantities
$(\frac{D}{4(D-1)})^{1/D}\alpha_j$ are equal to the averages of the Hubble
parameters
which in quantum theory we define as
$H_j\equiv d(\ln a_j)/dt$.
Since
 $\sum_{j=1}^D <H_j>=0$, we obtain that some dimensions have associated
positive constant Hubble parameters , that is they suffer a Quantum
Inflationary process, while the remaining dimensions must have associated
negative constant Hubble parameters , i.e. they suffer a Quantum Deflationary
process. The average of the total volume $<V>$ remains a constant. This phase
of the Universe we call \cite{qic1} the Quantum Inflation-Compactification
(QIC) era.

The above results follow also if instead of dust we introduce a massive
scalar field with its homogeneous degree of freedom described
quantum mechanically \cite{qic2}. This is a considerable improvement of
the model, since a perfect fluid description (of which dust is a particular
case) is only a
phenomenological approach while the description of matter as a scalar field
follows from first principles.

The appearance of a negative vacuum energy has been recognized as a
wide-spread property of many superstring models \cite{ramy}. Here we have
seen that a negative vacuum energy in the early Universe, which is usually
regarded as a disaster \cite{ramy} from a cosmological point of view, is in
fact a blessing,
since a  phase with negative vacuum energy can be the origin of a QIC era,
where the asymmetry between extra and ordinary dimensions was generated.

It is important to understand, at least at a qualitative level, how
this QIC era can evolve into a phase where the cosmological constant is
approximately zero and the dynamics is well approximated by the classical
theory. One scenario requires only the existence of a dilaton field with a
potential with
two local minima: one is an absolute  minimum with negative vacuum energy
density while the other has zero energy density. The QIC era is realized when
we consider oscillations of the scalar field around the absolute minimum. This
homogeneous degree of freedom behaves like dust, as mentio-ned above. The
explicit solutions \cite{qic2} show that the total scalar field energy (vacuum
plus coherent excitations) can be positive, so that a homogeneous tunneling to
the zero cosmological constant state is possible, since the volume of the
Universe is finite. Furthermore, at late times the phase with approximately
zero cosmological constant and growing volume can be stable against decay to
the negative cosmological constant state if a suitable condition on the scalar
field potential is satisfied \cite{cole} .



\begin{thebibliography}{99}


\bibitem{guth}
A. H. Guth, Phys. Rev., $\bf{D}$23 (1981) 347.

\bibitem{venus}
M. Gasperini, N. Sanchez, and G. Veneziano, Nucl.Phys. $\bf{B}$364 (1991) 365
and the many references to earlier works mentioned there.

\bibitem{dima}
A. Beloborodov, M. Demianski, P. Ivanov  and  A. Polnarev, Phys. Rev.
$\bf{D}$48 (1993) 503.

\bibitem{qic1}
E. I. Guendelman and A. B. Kaganovich, Int.J.Mod.Phys. $\bf{D}$2 (1993) 221.

\bibitem{early}
J. B. Hartle,  in {\em The Very Early Universe}, eds. G. W. Gibbons
and S. W. Hawking,  Cambridge University Press, Cambridge, 1983, p.72
and references  here.

\bibitem{kuchar}
K. V. Kuchar,  in {\em Proc. of the 4-th Canadian Conf. on
Gen. Rel. and Rel. Astrophys.}, eds. G. Kunstatter, D. V. Vincent
and J. G. Williams,  World Scientific, Singapore, 1992, p.211,
for a review of the problem of time in quantum gravity   .

\bibitem{qic2}
E. I. Guendelman and A. B. Kaganovich, Mod.Phys.Lett. $\bf{A}$9 (1994) 1141.

\bibitem{ramy}
R. Brustein and P. J. Steinhardt, Phys. Lett.$\bf{B}$302 (1993) 196.

\bibitem{cole}
S. Coleman and F. DeLuccia, Phys. Rev. $\bf{D}$21 (1980) 3305.






\end{thebibliography}
\end{document}